\documentstyle[12pt]{article}
\begin{document}
\baselineskip=20pt
\noindent
{\large\bf The Lense--Thirring Effect and Mach's Principle
}\\*[0.2cm]
\bigskip
\centerline{Hermann Bondi${}^\dagger$ and Joseph Samuel}
\centerline{Raman Research Institute, Bangalore 560 080,
India}
\bigskip
\bigskip
\begin{abstract}
We respond to a recent paper by Rindler on the 
``Anti--Machian'' nature of the Lense--Thirring effect.
We remark that his conclusion depends 
crucially on the particular formulation
of Mach's principle used.
\end{abstract}
\vspace*{6cm}
\bigskip
{$\dagger$ Permanent address: Churchill College, Cambridge, CB3 0DS,
England}
\newpage
\def\complex{{\rm I\!\!\!C}}
\def\real{{\rm I\!R}}
\section{Introduction}
In a recent paper, Rindler \cite{Rindler} has analysed the 
Lense--Thirring effect \cite{Lense,Chap,Brill} 
and concluded that the result is anti-Machian. 
Rindler uses a particular interpretation of Mach's principle. We
wish to stress here that Rindler's interpretation is only one amongst many. 
Indeed, the literature on this topic is so diffuse that we think
it desirable to set out a list
of interpretations that come to mind. Our list is far from 
exhaustive, but it is long enough to make numbering different
versions necessary.

We begin with Mach0, which is 
the basis of the whole idea:
{\it The universe, as represented by the 
average motion of distant galaxies \cite{Fo1} does not 
appear to rotate relative to local inertial frames.}

We illustrate this point by a modern version of Newton's famous 
bucket experiment: the Sagnac effect. This effect provides an operational
method for an observer to decide, by local measurements, if she is rotating.
Consider an astronaut in an enclosed spaceship
with angular velocity $\omega$.
The astronaut takes a closed circular fibre 
optic tube at rest with respect to the spaceship
and sends two rays of monochromatic
laser light in opposite directions around the tube. 
These rays are made to interfere \cite{Fo2}  
after each ray has gone round once.
If the spaceship is rotating, the corotating ray will take longer
to come around than the counter--rotating one, leading to an
arrival time difference, which can be observed as a fringe
shift. The time difference is given by:
$\Delta t=-4A\omega/c^2$, where $\omega$ is the angular velocity
of the spaceship 
and $A$ is the area enclosed by the tube. 
Using the Sagnac effect, one can
by experiments internal to the spaceship, so arrange the angular
velocity of the spaceship that the Sagnac shift (defined
as $\Delta t/2$) vanishes. A frame at rest
with respect to such a spaceship is called a
locally non rotating frame. 
Sit
in this frame, look
up at the sky and note that the distant galaxies 
are still. Mach's principle
(Mach0) is the experimental observation that the inertial frame
defined by
local physics (zero Sagnac shift) coincides with the frame in
which the distant objects are at rest. 

Mach0 is an experimental observation and not a principle. 
One could interpret Mach's writings as a suggestion to 
construct a theory in which Mach0
appears as a natural consequence. But Mach's writings have been variously
interpreted. Our purpose here is to list a number of
interpretations of Mach's principle and view them in the light
of currently accepted theories in an effort to refine and clarify the idea.

We do have at our disposal
two well established theories of space,
time, gravity, matter and motion 
--Newton's and Einstein's--
both experimentally succesfull in their respective domains of validity.
Newton's holds that space and time are absolute. Einstein's
holds that space time geometry is affected by matter.
There is no question (as Rindler observes) that these experimentally
successful theories are here to stay regardless of whether they 
satisfy any of the rather philosophical criteria embodied in
Mach's principle.

\section{Versions of Mach's Principle}
Recent discussions of Mach's Principle, including this one, have
greatly benefitted from the 1993 Conference organised by J.
Barbour and H. Pfister and the excellent book \cite{Book}
resulting from it. A glance at the book (note especially J.
Barbour's list on page 530) will show that there
have been numerous interpretations of Mach's writings. 
For an authoritative account of the history of Machian ideas,
the reader is referred to \cite{Book}. 

We now list a few versions of Mach's principle which appear in
the literature. Each statement of Mach's principle, will be accompanied by a
declaration of the theoretical
framework in which it is intended to apply. Two levels of compatibility
will be considered: Does the particular statement of Mach's
Principle make sense in the theory, and secondly, is it
satisfied by it?
We use the letters N and E to refer to Newtonian and
Einsteinian space time.
Even within
Einstein's theory there is a further dichotomy-- is one discussing
cosmology 
(the whole universe) or an isolated system embedded in an asymptotically
flat space time? This distinction is made by the notation EA for
asyptotically flat spacetimes and EC for relativistic 
Cosmologies. Our purpose
in compiling this list is to draw attention to the diversity of
ideas that pass under the guise of ``Mach's principle''.
(Page numbers refer to \cite{Book} unless otherwise indicated.)

\begin{itemize}
\item Mach1: {\it Newton's gravitational constant $G$ is a dynamical field.}
(Makes sense in N, EA, EC.) Mach1 is not true in N or E.
This version applied to Einstein's theory 
has led to Brans--Dicke Theory\cite{Brans,Wein}. 

\item Mach2: {\it An isolated body in otherwise empty space has
no inertia (pp 11,39,181, 185).}  
(Makes sense in N, EA, EC.)
 Neither Newtonian nor Einsteinian gravity satisfy this version.
In both theories the motion of an isolated body is determined 
and not arbitrary.
\item Mach3: {\it local inertial frames are affected by
      the cosmic motion and distribution of matter (p92).} 
(Makes sense in N, EA,
EC \cite{Fo3} .) This
version is closest to the bucket experiment. In this form,
Newton's theory is in clear conflict with Mach3. 
Einstein's theory is not (see section 4 below).
\item Mach4: {\it The universe is spatially Closed (p 79).}
(Makes sense only in EC.) We do not know if Mach4 is true.
\item Mach5: {\it the total energy, angular and linear momentum of
the universe are zero (p237).}(Makes sense in N, EA, EC.) It is not
true in N and EA. In EC it is claimed  \cite{King} 
that the total angular momentum
of a closed universe must vanish.

\item Mach6: {\it Inertial mass is affected by the global distribution
of matter (pp 91,249).} Makes sense in (N, EA, EC). Is not true 
in any of them.
Hoyle and Narlikar \cite{Hoyle} proposed a theory in which implements Mach6.

\item Mach7: {\it If you take away all matter, there is no more
space \cite{Edd}.} Makes sense in (N,EA,EC). Not true in any of them.

\item Mach8: $\Omega= 4 \pi \rho G   T^{2}$ {\it is a definite number
of order unity (p475).}
(Here, $\rho$ is the mean density matter in the universe and $T$
is the Hubble time. Makes sense in EC only.) $\Omega $ 
does seem to be of order unity
in our present universe, but note that of all EC models, only
the Einstein--DeSitter makes this number a constant, if $\Omega$
is not {\it exactly} one.
Making a theory in  which this approximate equality appears natural
is a worthwhile and ongoing effort (eg inflationary cosmologies).

\item Mach9: {\it The theory contains no absolute elements
(\cite{Ein}.} (Makes sense in N, EA and EC)
This version is clearly explained by J\"urgen Ehlers
in \cite{Book} p 458. The elements (fields, for example) 
appearing in the theory can be
divided into dynamical (those that are varied in an Action principle)
and absolute (those that are not). The Action principle leads to
equations for the dynamical fields to satisfy. The absolute elements
are predetermined and unaffected by the dynamics.

Newton's theory does not satisfy Mach9 (space and time are absolute)
and neither does EA
(asymptotic flatness introduces an absolute element--the 
flat metric at infinity). EC does satisfy Mach9 \cite{Shu}.
From the point of view of invariance groups (J.L Anderson, A. Trautmann,
quoted on p 468 \cite{Book})
Mach9 is the requirement that the invariance group of the theory
is the {\it entire} diffeomorphism group of spacetime. Viewed in
this light Mach9 is just the principle of general covariance.

\item Mach10: {\it Overall rigid rotations and translations of a system
are unobservable.} (This version makes sense {\it only} in N; 
In Einsteinian spacetime 
one has no idea what a rigid rotation is anymore than one knows what a
rigid body is.) This is {\it not} satisfied
in Newtonian theory. 
If one insists on the principle and
constructs a theory which satisfies it,
one is led \cite{Barbour} to a 
class of models (called ``relational'' by Barbour and
Bertotti \cite{Barbour}). There is considerable literature
on these models \cite{Book,Don}.
We spend a few words on these models and their
connection with Newonian theory.

{\it Relational Models:} Let $x^i_a$ , $i=1,2,3$, $a=1...N$ be
the positions of N particles in Newtonian spacetime and
$p_{ia}$ their conjugate momenta. The Hamiltonian $H(x,p)$
determines the time evolution of $(x^i_a,p_{ia})$ via Hamilton's equations.
The transformation 
\begin{eqnarray}
x^i_a(t)&\rightarrow &R^i_j(t) x^j_a(t) \nonumber\\
p_{ia}(t)&\rightarrow &R^j_i(t) p_{ja}(t),
\label{trans}
\end{eqnarray}
where $R^i_j(t)$ is an arbitrary time dependent rotation 
matrix maintains the
distance relations between the $N$ particles. If a model is relational
\cite{Barbour}, such a transformation is unobservable,
like a ``gauge transformation'' in electrodynamics. From Dirac's
theory of constrained systems \cite{Dirac,NM}, 
it follows that the transformations
(\ref{trans}) must be generated by first class constraints. The generator
of overall rotations of the system is the total angular momentum:
$$
J^i=\Sigma_a\epsilon^{ijk}x_{aj} p_{ak}.$$
Thus the system is subject to the constraints
$$\phi^i(x,p):=J^i-C^i\approx 0,$$
where $C^i$ are constants. The requirement that the constraints
be first class in the sense of Dirac \cite{Dirac} forces the constants
$C^i$ to vanish.

The extended Hamiltonian in the sense of Dirac is 
$$H_E(x,p)=H(x,p)+\omega_i J^i,$$
where $\omega^i$ are arbitrary functions. 
While we have only dealt with overall rotations in (\ref{trans}),
one can similarly deal with arbitrary translations and arbitrary
time reparametrizations.
Relational models can be thus derived from Newtonian Hamiltonian
mechanics by imposing constraints on the phase space so
that the total angular momentum, momentum and Energy vanish. 

These relational models are clearly
distinct from Newtonian theory. For instance, Newtonian theory admits
solutions with nonzero angular momentum (like the solar system
in an otherwise empty universe) while relational models do not
permit such solutions. 

\section{Rindler's Criticism}
We now briefly summarise Rindler's argument. Consider the earth
in an otherwise empty universe. Let $O$ be a
reference frame rigidly attached to the earth.
Suppose that a gyroscope G  is taken
around the earth in the equatorial plane 
along a circle of radius $r$ with a constant clockwise
angular velocity $\Omega$. To keep track of orientations, we suppose 
the earth and the gyroscope marked with cross hairs (as in Fig.1
of Rindler). We arrange that the orientation 
of G relative to the earth's is constant during
the motion. (Rindler uses the Schwarzschild metric outside the earth
to compute $\alpha$ the precession rate of the gyroscope. We
choose the radius $r$ to set $\alpha $ to zero. It simplifies the argument.)

Now view the situation from the point of view of an observer
$O'$, who rotates rigidly relative to $O$ with constant clockwise
angular velocity $\Omega$. $O'$ sees the earth rotating anticlockwise
with angular velocity $\Omega$, the centre of the gyroscope at rest.
Notice however, that the gyroscope (which was not rotating with
respect to $O$) now rotates anticlockwise with angular velocity 
$\Omega$ relative to $O'$. Thus the gyroscope rotates {\it in
the same sense} as the earth. 

It follows from Mach10 that 
a rotating body in otherwise empty space
makes the local compass of inertia take up {\it all}
of the body's angular velocity. Applied to the earth, which is not in
empty space but in the universe, one would expect that the 
effect of the earth on the gyroscope should be considerably
diluted by the effect of the rest of the universe. Thus one would expect
that the local compass of inertia would take up a small
{\it positive} fraction of the earth's angular velocity. The sign of this
effect is everywhere positive unlike the sign of the
Lense--Thirring effect. This is the basis for Rindler's
conclusion that the Lense--Thirring effect is Anti--Machian.

\end{itemize}

\section{The Lense-Thirring effect as Machian}

We now show that one can arrive at the opposite conclusion from
Rindler's by using a different version of Mach's Principle.
We use the often employed exact analogy between
rotation in General Relativity and magnetic fields \cite{Bala}
to deduce that the slight
influence of a spinning body on the rotation of the near-by compass
of inertia goes with that of the body near the poles and in the opposite
sense in the equatorial plane. 

{\it The Lense--Thirring effect:} Consider a stationary spacetime
{\it i.e} one with a timelike Killing vector $\xi$:\,$\nabla_a\xi_b
+\nabla_b\xi_a=0$. One can adapt the time coordinate to $\xi$ so
that $\xi=\partial/{\partial t}$ and the metric assumes the form:
$$ds^2=g_{00}(dt+A_i dx^i)^2-\gamma_{ij}dx^idx^j\label{metric},$$
where $A^i=g_{0i}/g_{00}$.
The coordinate transformations that preserve this form include
$t\rightarrow t+\alpha(x^i)$, which physically represents the resetting
of clocks. Under such transformations $A_i$ transforms as $A_i\rightarrow
A_i+\nabla_i \alpha$ like the vector potential in
electrodynamics. Consequently its curl $F_{ij}:=\partial_i A_j
-\partial_j A_i$ is invariant and represents rotation of the spacetime
(more geometrically, the failure of $\xi $ to be hypersurface orthogonal).
It is easily seen that a stationary Sagnac tube will measure a
Sagnac shift of $\oint A_i dx^i$. 
A locally nonrotating
Sagnac tube (one that measures zero Sagnac shift) 
will appear to rotate as viewed from infinity.
The angular velocity of rotation has the spatial distribution of
a dipole magnetic field and {\it reverses sign between the equator
and the poles}. As we show below this is {\it exactly} what one expects
from Mach's principle (Mach3).

If one applies Mach's Principle in the form Mach3 to understanding
rotation in General Relativity, one sees that the prediction of 
Mach3 agrees with
the sense of the Lense--Thirring effect. 
If one is stationary at
the north pole of the earth one sees the earth rotating anticlockwise
and one also sees a ``non rotating'' gyroscope (one which
registers a null Sagnac effect) rotating anticlockwise.
(The magnitude of the effect is not in question here
only its sign.) The agreement between the sense of the
Lense--Thirring effect and Mach3 also
extends to the equatorial plane. 
It is true that the sense of the Lense--Thirring effect reverses
at the equator.
It is also true that the prediction of Mach3 reverses: An
observer in the equatorial plane sees the {\it nearer parts} of the Earth
moving past her sky in an {\it clockwise} direction around the
North Star. While the
{\it further parts} are moving in an {\it anticlockwise} direction, 
the sense
of the effect is dominated by the nearer mass. The net effect is
as {\it clockwise} rotation of a locally nonrotating gyroscope.
Thus Mach3
agrees with the Lense--Thirring effect both at the poles and the
equator. (For a somewhat different argument leading to the same result,
see the articles by Schiff and Thorne, quoted on page 321 of \cite{Book}).

\section{Conclusion}
The list given above shows the variety of interpretations that Mach's
writings have spawned. Some of them express the idea 
``Cosmic conditions affect local physics''. Others state requirements to
be satisfied by physical theories. There are also logical
relations between some of the versions: for instance Mach10
(which is formulated in N) implies that the total angular momentum,
momentum and energy
of the Universe is zero. This is precisely the content of Mach5,
which is formulated more generally. On the other hand, Mach1 has
no obvious connection with Mach0.

To us, the most remarkable feature of the list (which Rindler's paper
\cite{Rindler} brings to light) is that two
entries in it (Mach3 and Mach10) give rise to {\it diametrically
opposite} predictions, when applied to a simple physical situation.
By popular usage Mach's principle has acquired a range of meanings,
some of which are in conflict with each other. Mach's writings
have been a source of inspiration to many (including Einstein).
We hope that our effort at distinguishing between existing
versions of Mach's Principle will serve to clarify ideas and eliminate
needless controversy.

{\it Acknowledgements}: It is a pleasure to thank G.W. Kang,
B.R. Iyer, H. Pfister, W. Rindler and C.S. Shukre for their comments on an
earlier version of this manuscript. HB thanks the Indian Academy
of Sciences, Bangalore as this paper arose during his tenure
as Raman Professor of the Academy.


\begin{thebibliography}{99}
\bibitem{Fo1}
Though the early authors spoke of the
``fixed'' stars, their movement in our rotating Galaxy is indeed
observable with modern instrumentation.
\bibitem{Fo3}
There is some vagueness in
this formulation as applied to Einstein's theory: Does
``matter'' include the gravitational degrees of freedom? 
We restrict attention to
stationary spacetimes to eliminate
gravitational waves.
\end{thebibliography}
\end{document}